\documentclass[12pt]{article}
\newcommand{\GeV}{\textrm{ GeV}}

\newcommand{\tb}{\tan \beta}
\newcommand{\TeV}{\textrm{ TeV}}
\usepackage{amsmath}\usepackage{amssymb}
\usepackage{latexsym}
\usepackage[dvips]{graphicx}
\usepackage{pstricks}
\usepackage{bm}
\usepackage{axodraw}
\newcommand{\be}{\begin{equation}}
\newcommand{\ee}{\end{equation}}

\def\beq{\begin{equation}}
\def\eeq{\end{equation}}

\def\al{\alpha}
\def\bt{\beta}

\def\ga{\gamma}
\def\de{\delta}

\def\si{\sigma}

\def\La{\Lambda}
\def\lam{\lambda}

\def\sq{\sqrt}

\def\l{\left (}
\def\r{\right )}

\def\fr{\frac}
\def\la{\label}
\def\hs{\hspace}
\def\vs{\vspace}

\begin{document}

\begin{flushright}
OSU-HEP-08-04\\
FERMILAB-PUB-08-151-T\\
June 10, 2008 \\
\end{flushright}

\vs{1.5cm}

\begin{center}

{\Large\bf
Perturbativity and a Fourth Generation in the MSSM}

\end{center}
\vspace{0.4cm}
\begin{center}
\renewcommand{\thefootnote}{\fnsymbol{footnote}}
{\large {}~Zeke Murdock ${}^{1,2}$\footnote{Scientific Visitor at Fermilab, E-mail:
zekemurdock@gmail.com},{}~ {}~S.~Nandi ${}^{1,2}$\footnote{Scientific Visitor at Fermilab, E-mail: s.nandi@okstate.edu},{}~ {}~Zurab
Tavartkiladze ${}^{1}$\footnote{E-mail:
zurab.tavartkiladze@okstate.edu} } \vspace{0.5cm}
\\
{\small ${}^{1)}${ \em{Department of Physics and Oklahoma Center for
High Energy
Physics,}}}\\
Oklahoma State University, Stillwater, OK 74078, USA,\\
{\small${}^{2)}${ \em{Fermi National Accelerator Laboratory, P.O. BOX
500, Batavia, IL 60510, USA}}}

\end{center}
\vspace{0.6cm}

\begin{abstract}
    \noindent
We study an extension of the MSSM with a fourth generation of chiral matter.
With this extension no value of $\tb$ allows the theory to stay perturbative up to the GUT scale. We suggest one model with extra
vector-like states at the TeV scale that allows perturbativity all the way up to the GUT scale.

\end{abstract}

\section{Introduction}
\label{intro}

The repetition of the quark-lepton families is one of the great mysteries of particle physics. Despite its great success in describing the nature of strong and electroweak (EW) interactions, the Standard Model (SM) does not predict the number of families. What is the principle limiting the number of chiral families? Why not have a fourth generation or even more? The masses of the three observed families have a strong hierarchical pattern. Only the top quark mass ($m_t\simeq 172.6$~GeV) lies close to the EW symmetry breaking scale. This, within the SM, suggests the Yukawa coupling of the top quark should be $\lam_t\simeq 1$. All remaining Yukawa couplings are suppressed. Thus, with only three observed families,  $\lam_t$ and the three gauge couplings $g_{1,2,3}$ would play an essential role in dynamics upon performing renormalization group (RG) studies.  The situation may be modified within a two Higgs doublet SM and MSSM. In these models, due to the parameter $\tan \bt =v_u/v_d$ (the ratio of the VEVs of the up type to the down type Higgses) $\lam_b$ and $\lam_{\tau }$ can also be large ($\sim 1$ for $\tan \bt \approx 60$). How would the picture change if there were a fourth family?

  Current lower limits on the masses of the 4th generation fermions at $95\%$ C.L.
   are \cite{pdg}:
  \beq
   m_{t'}\geq 220~{\rm GeV}~,~~~ m_{b'}\geq 190~{\rm GeV}~,~~~
   m_{\tau'}\geq 100~{\rm GeV}~, ~~~ m_{\nu'}\geq 50~{\rm GeV}~.
  \la{bounds-4-genmas}
  \eeq
When these masses are translated to the values of their Yukawa couplings, we find the possibility of couplings larger than $\lam_t$. Moreover, the bound on $m_{\nu'}$ indicates the existence of at least one massive neutrino with mass near the EW scale.

Due to the possible existence of large new Yukawa couplings, a study should be performed and the validity of the perturbative treatment must be examined. As we will show, within the MSSM with a 4th family, there is no value of $\tan \bt $ that allows the perturbativity of the couplings up to the GUT scale. This fact suggests a lower cutoff scale.  If there is such a cutoff scale, it should be related to new physics which take care of the self consistent ultraviolet (UV) completion. Can such a completion be constructed? A positive answer would be encouraging for model building as well as for further investigations with various phenomenological implications.

Even without focusing on UV completion of the theory, any extension of the SM or MSSM should be in accord with low energy observables. Some previous works  \cite{Frampton:1999xi}-\cite{tait} have discussed the effect of a 4th generation on the EW precision parameters $S$, $T$. These constrain the masses of $t'$ and $b'$ quarks.  In agreement with them, we find the effect on the $U$ parameter is well within the 3$\sigma$ limits of PDG. 

\begin{figure}[!ht]
\begin{center}
      \resizebox{90mm}{!}{\includegraphics{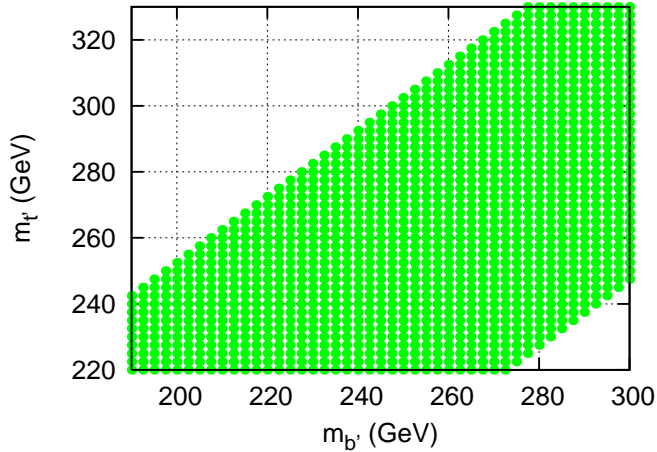}}
\end{center}
 \caption{ Plotting allowed quark masses using $3\sigma$ limits of S and T.   Here, $M_h=115 \GeV$, $m_{\tau'}=150 \GeV$, and $m_{\nu'}=100 \GeV$. $(\Delta S,\ \Delta T)$ for the leptons are $(0.008,\ 0.045)$.  The contribution to $U$ is negligible.}
\label{fig:both}
\end{figure}

Assuming that the mixings of the fourth family matter with the observed three generations are minimal, most of the constraints come from the self energy diagrams of $W^\pm$ and $Z^0$ gauge bosons.
In Ref. \cite{tait} it was found that with
$        m_{t'}-m_{b'} \simeq \left(1 + \frac{1}{5} \ln \fr{M_h}{115\GeV}
        \right) \times 50\GeV$,
the new contributions to the parameters $S$ and $T$ get minimized. In particular, with $M_h = 115 \GeV$ one obtains $m_{t'}-m_{b'}\simeq 50\GeV$. Using analytical expressions given in Ref. \cite{he} and the experimentally allowed ranges of $S$, $T$, and $U$ at $1\si $ \cite{pdg}:
 \begin{eqnarray}
        S & = & -0.13 \pm 0.10~,\nonumber \\
        T & = & -0.13 \pm 0.11~,\nonumber \\
        U & = & 0.20 \pm 0.12~,
\label{stu}
    \end{eqnarray}
we can  derive further constraints on $m_{t'}$ and $m_{b'}$. In Fig. \ref{fig:both} we show the allowed regions for $m_{t'}$ and $m_{b'}$. For these analysis we have allowed $3\si$ deviations in Eq. (\ref{stu}).

A fourth generation of chiral matter would also affect the Higgs sector. This will give more interesting insights \cite{Arik:2001iw, fok} within a SUSY framework. As is well known, in MSSM the value $\tan \bt \approx 1$ is disfavored due to the LEP lower bound on a lightest CP even Higgs boson mass $M_h\geq 114.4$~GeV. In the MSSM, at tree level  $M_h^2 = M_Z^2\cos^2 {2\beta}$. Taking $\tan \bt \simeq 1$, the tree level mass vanishes.  Loop corrections are not sufficient to raise $M_h$. When a 4th generation is added, the situation is even more drastic because in order to preserve perturbativity $\tan \bt $ cannot be much greater than 1. This is an additional motivation for new physics.

This leads us to believe that the MSSM with a 4th family should be extended further. In this paper we suggest one such extension with vector like states having masses at the TeV scale. As an outcome of the proposed model, we obtain perturbativity of the couplings all the way up to the GUT scale with $\tan \bt \sim 2$. This avoids the difficulties discussed above, and is promising for the possibility of embedding the whole scenario in a grand unified theory.

The paper is organized as follows. In section \ref{theory} we discuss theoretical bounds: problems arising from perturbativity considerations that limit $\tb$ and implications on Higgs physics.  In section \ref{model} we present our model which allows perturbativity of all couplings up to the GUT scale and extends $\tan \bt $  up to $\sim 2$ such that the LEP bound on $M_h$ can easily be satisfied. The model has extra vector-like states which can be detected at the LHC. The summary of our work and conclusions are presented in section \ref{conc}.

\section{Theoretical Bounds and Some Implications}
\label{theory}

In this section we discuss bounds coming from theoretical
considerations and discuss some implications of theories with a
new heavy chiral fermion family.

\subsection{Bounds from Tree Level Unitarity}

The upper bound on a heavy chiral fermion's mass comes from the unitarity of scattering amplitudes. We assume that fermion mass is generated through the Yukawa coupling of the fermion with a fundamental Higgs doublet. In this case, for the heavy quark doublet $Q$ with mass $m_Q$ the
 $Q\bar Q\to Q\bar Q$ scattering
$J=0$ partial wave amplitude at tree level (at energies $\sq{s}\gg m_Q$) is given
 by \cite{Chanowitz:1978uj}:
\beq
|a_0|\approx \fr{5}{4\sq{2}\pi }G_Fm_Q^2~,
\la{a0Q}
\eeq
and the unitarity requirement $|a_0|<1$ gives the upper bound
\beq
m_Q^2<\fr{4\sq{2}\pi }{5G_F}\simeq (552 \GeV)^2~,
\la{upbQ}
\eeq
as was first obtained in \cite{Chanowitz:1978uj}. The analogous bound for the leptonic
 doublet $L$
\beq
m_L^2<\fr{4\sq{2}\pi }{G_F}\simeq (1.23 \TeV)^2~,
\la{upbL}
\eeq
is higher. As we see, the current experimental direct bounds in Eq. (\ref{bounds-4-genmas})
 are not in conflict with the theoretical
upper bounds of (\ref{upbQ}) and (\ref{upbL}) derived at tree level. As we discuss below, the inclusion of loop corrections and the requirement of perturbativity will imply stringent theoretical bounds on Yukawa couplings.

\subsection{Bounds from Perturbative RGE}

Here we focus on MSSM with a 4th generation. The reason
for the SUSY framework is twofold. First of all, low scale SUSY
is the most appealing extension of SM in order to solve the gauge
hierarchy problem. Second, as it turns out, more stringent bounds are
obtained in the SUSY setup and for demonstrative purposes it is most
useful.
 The discussed mechanisms (presented in the next section)
for solving various problems could be also applied for SM and on two Higgs doublet SM.

\begin{figure}[ht]
   \centering
        \includegraphics[scale=1.7]{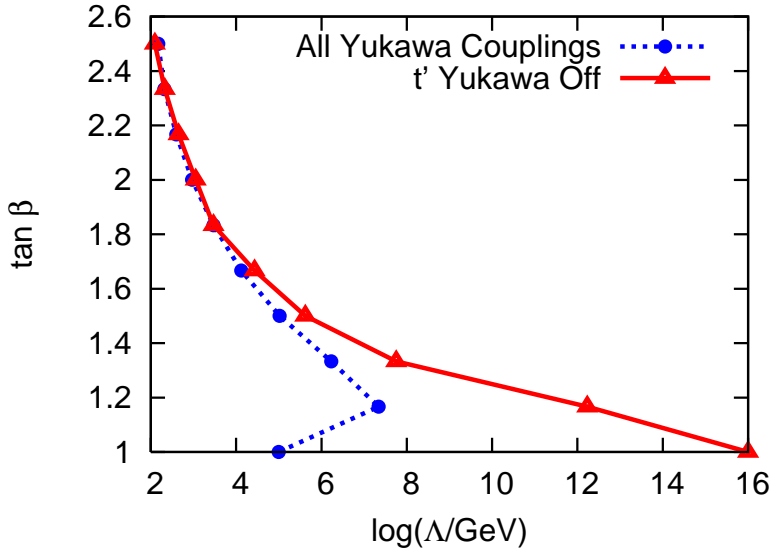}
    \caption{Plotting $\tb$ vs. $\Lambda$, the scale at which one of the Yukawa couplings becomes non-perturbative.  
    Dotted line corresponds case with all Yukawa couplings taken into account.  For masses we took lowest allowed
    values from Eq. (\ref{bounds-4-genmas}).}
    \label{fig:tanbeta}
\end{figure}

The superpotential couplings involving 4th generation matter superfileds are
\beq
W_4=\lam_{t'}q_4u^c_4h_u+\lam_{b'}q_4d^c_4h_d+\lam_{\tau '}l_4e^c_4h_d+
\lam_{\nu_{\tau'}}l_4Nh_u,
\la{4th-superpot}
\eeq
where $N$ is a right handed neutrino (complete singlet of MSSM) responsible for
the Dirac mass generation of $\nu_{\tau'}$.
Yukawa couplings defined at corresponding mass scales can be expressed as
$$
\lam_{t'}(m_{t'})=\fr{m_{t'}}{|1+\de_{t'}|v\sin \bt }~,~~~~~~~~~
\lambda_{b'}(m_{b'})=\frac{m_{b'}}{|1+\de_{b'}|v\cos \bt }~,
$$
\beq
\lambda_{\tau'}(m_{\tau'})=\frac{m_{\tau'}}{|1+\de_{\tau'}|v\cos \bt
}~,~~~~~~~~~
\lam_{\nu_{\tau'}}(m_{\nu_{\tau'}})=\fr{m_{\nu_{\tau'}}}{|1+\de_{\nu_{\tau'}}|v\sin
\bt }~, \la{lambda-t1b1l1n1} \eeq where $\de_{\al }$ ($\al={t'},
{b'}, {\tau'}, {\nu_{\tau'}}$) exhibit the 1-loop finite corrections
emerging after SUSY breaking \cite{Hall:1993gn}. Since we are
dealing with large Yukawa couplings($\sim 2$), these corrections can
be as large as $25\%$ and should be taken into account. For
examining the RG perturbativity, one should take the values for masses satisfying the bounds in Eq. (\ref{bounds-4-genmas}) and run each Yukawa coupling from the
corresponding mass scales up to higher scales. In Ref.
\cite{gunion} this analysis was done with the fourth generation fermion
masses smaller than the top mass. This was in accord with the
experimental bounds that existed at that time.   They found that if
$\tb<3$ all Yukawa couplings could be perturbative up to the GUT
scale.  Given the current lower bounds on quark and lepton masses,
we find this is no longer the case. When one uses the
renormalization group equations for  evolving the Yukawa couplings
from low scale up to higher energy scales, the couplings rapidly grow
and blow up. For example for $\tb=2$, $\lambda_{b'}$ becomes
non-perturbative at about $1$~TeV.  As $\tb$ increases, it is more
difficult to tame the Yukawa coupling. This is shown in Fig.
\ref{fig:tanbeta}.

We have assumed the validity of perturbative RG for Yukawa couplings $<2.5$.
For these analysis we ignored all $\de_{\al }$s (i.e. set $\de_{\al }=0$, keeping in mind that unknown soft breaking terms allow more flexibility),
however even the values $\de_a\sim 1/4 $ do not change the situation much.

  It is clear from this figure that \emph{no} value of $\tb$ allows perturbative
  calculation all the way up to the GUT scale.
Perturbativity puts a strict upper bound on the mass of the $b'$ quark.  For $\tb = 1.5$
we calculate this limit to be about $\approx 100\GeV$.  This value is below the experimental lower bound of $190 \GeV$.  If a fourth generation exists, this provides a strong reason to introduce new physics at the TeV scale.
In order for this to work, the cutoff scale of the theory should be near the TeV scale.  Without any UV completion we have a strongly coupled theory at the TeV
scale. What are the solutions to this problem? In section \ref{model}
we will introduce a specific model with  new physics at the TeV
scale that will allow values of $\tb$ up to $\sim 2$ with perturbativity all the way up to the GUT scale $\approx 2\cdot 10^{16} \GeV$.

\subsection{Implications for Higgs Physics}

In the MSSM with large $\tan \bt $ the lightest Higgs boson mass has an upper bound $M_h\stackrel{<}{_\sim }125$~GeV.  Even if $\tb$ is large, the mass at tree level can be no larger than $M_Z$.  This is an even bigger problem when one introduces a fourth family.  The new quarks limit $\tb$ to small values, thus reducing the tree level contribution for the lightest Higgs mass. Luckily at the same time they provide additional loop corrections to the lightest Higgs mass.  The one-loop top-stop radiative corrections to the Higgs mass squared can be
simplified as:
\begin{equation}
\Delta(M_h^2) \simeq \frac{3}{4\pi^2}\fr{m_t^4}{v^2} \ln \fr{m_{\widetilde{t_1}}
 m_{\widetilde{t_2}}}{m_t^2}~.
\label{loop}
\end{equation}
The new $t'$ and $b'$ quarks and their superpartners will also contribute to the Higgs mass.
These corrections can enhance the Higgs mass \cite{Haber:1990aw,Ellis:1990nz}.
  When $\tb>1$,
the correction from the $b'$ quark has a similar form, but it is negative.
  If $m_{b'}>m_{t'}$ then there is a problem, as the overall correction will be
  negative.  When $m_{t'}>m_{b'}$, with constrained mass splitting displayed
  in Fig. \ref{fig:both}, there is still a sizable
  positive correction of about $(60\GeV)^2$.  With $\tb \sim 2$ this puts an upper bound,
  $M_h\stackrel{<}{_\sim }130\GeV$, greater than the LEP lower bound of $114\GeV$.

  The existence of a 4th chiral family in the mass range of
$(200-300)\GeV$ will have a significant impact on the Higgs signals at the
  LHC \cite{tait}. The most dominant production mechanism for the light Higgs
  boson is its production from gluon-gluon fusion via a top
  quark loop \cite{Georgi:1977gs}. With the 4th chiral family, there will be additional
  contributions from the non-degenerate $t'$ and $b'$ loops. Thus the Higgs
  productions will be significantly enhanced. Also, for the light
  Higgs with mass below ~$130\GeV$, the Higgs decaying to two photons
  is the most clean channel for detection at the LHC. With the
  additional contributions from the $t'$ and $b'$ quarks in the loops, the
  two photon branching ratio will also be enhanced. The other
  possible mode for the light Higgs detection is the $t
  \overline{t}h$ mode, and the subsequent decay of the Higgs to
  $b \overline{b}$. This mode has been downgraded by recent studies
  mainly due to low production rate and large SM background.
  However, with the 4th family quarks, there will be additional
  contributions to the Higgs production via the $t' \overline{t'} h$
  and $b' \overline{b'} h$ modes. Thus the Higgs detection via this
  channel may become viable.

\section{The Model with Perturbative UV Completion}
\label{model}

If the LHC discovers a fourth chiral family, it will be a great challenge for theorists to
build self consistent models. There are several reasons for this. First of all, from
existing experimental bounds it follows that the Yukawa couplings for $t'$ and $b'$
should be large.
Let us be more specific. If the theory is one Higgs doublet Standard Model (SM), then
 the bounds
$m_{t'}\geq 220$~GeV and $m_{b'}\geq 190$~GeV imply that near these mass scales we have
 $\lam_{t'}\geq 1.26$ and $\lam_{b'}\geq 1.1$.
The situation is more drastic within the MSSM. The above bound for the $m_{b'}$ gives
 $\lam_{b'}\geq 1.1\sq{1+\tan^2\bt }$ which for $\tan \bt \simeq 3$ gives
  $\lam_{b'}(m_{b'})\geq  3.45$, a non-perturbative value. Therefore, the (tree level)
   perturbativity
suggests the upper bound  $\tan \bt \leq 2.5$. However, as we saw in the previous section,
 after taking into account RGE
effects, the requirement of perturbativity up to higher scales
prefers even lower ($\stackrel{<}{_\sim }1.5$) values of  $\tan \bt
$. This may lead to clash with the LEP bound on the lightest Higgs
boson mass $M_h\geq 114$~GeV.
For $\tan \bt \sim 1$, in MSSM with three families it is difficult to satisfy this
 bound. As RGE studies discussed in section \ref{theory} show, no value of $\tan \beta$
  allows perturbativity up to the GUT scale for $M_{\rm GUT}\simeq 2\cdot 10^{16}$~GeV.
  What are the possibilities to overcome these difficulties? The solution is some
  reasonable extension which modifies RG running above the TeV scale. Here we suggest
   one simple extension which allows perturbativity up to the  $M_{\rm GUT}$ with less
    constraint on $\tan \bt $.

Our proposal is the following. The couplings $\lam_{t'}$, $\lam_{b'}$ and $\lam_{\tau '}$
 are derived quantities in a low energy effective theory.
They are generated after decoupling of additional vector like states with mass
 $\La_4\sim {\rm few}\cdot $TeV. Above $\La_4$, new interactions
appear in the RGE and this makes the theory perturbative all the way up
to $M_{\rm GUT}$. We discuss the realization of this idea within the
framework of the MSSM, however,
 non-SUSY models can be constructed with equal success.

We introduce two additional vector like pairs $(H_u+H_d),\
({H_u}'+{H_d}')$ of Higgs superfields, where   $H_u,{H_u}'$ and
$H_d,{H_d}'$ have the same quantum numbers
 under the MSSM gauge group as the up type $(h_u)$ and the down type $(h_d)$ Higgs
  superfields.
 These $H$-states are accompanied by two pairs of vector like quarks $(D^c+\bar D^c)$,
 $({D'}^c+\bar D^{'c})$, where $D^c$ has the quantum numbers of the down type quark $d^c$.
 Introduction of $D$-states are suggestive: they, together with $H$-states, effectively
 constitute complete $SU(5)$ multiplets and therefore gauge coupling unification
 can be maintained at 1-loop approximation.

\begin{figure}[!ht]
   \begin{flushleft}
       \hs{1cm}\includegraphics[scale=1]{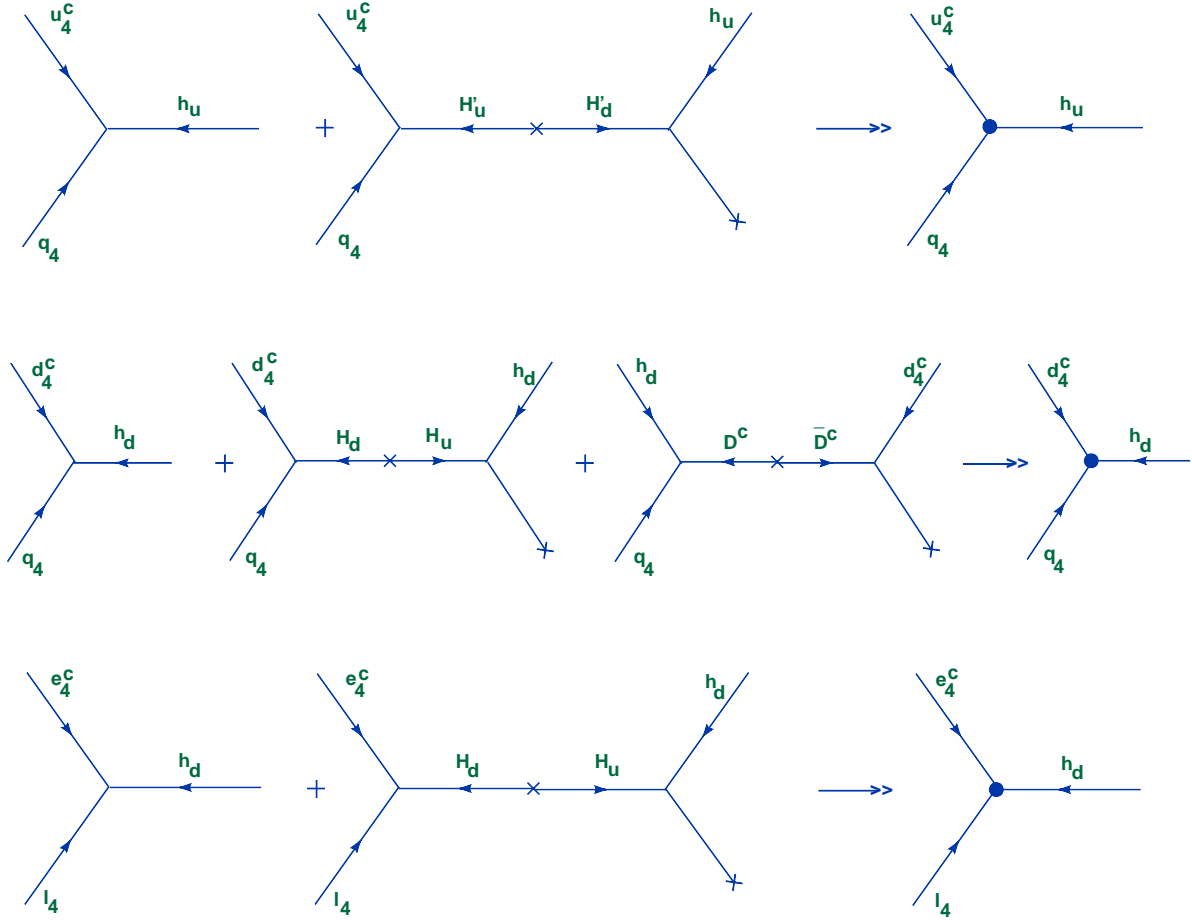}
        \caption{Diagrams generating Yukawa couplings $\lam_{t'}$, $\lam_{b'}$ and $\lam_{\tau'}$.}
        \label{fig:4seesaw}
   \end{flushleft}
\end{figure}

We will consider the following superpotential couplings
\begin{eqnarray*}
W_{4}&=&\lam_{t'}^{(1)}q_4u^c_4h_u+\lam_Uq_4u^c_4H_u+\lam_{b'}^{(1)}q_4d^c_4h_d
+\lam_Dq_4d^c_4H_d+\lam_D' q_4D^ch_d+\lam_{\tau '}^{(1)}l_4e^c_4h_d\\
     & &+\lam_El_4e^c_4H_d-M_HH_uH_d-M_{H'}{H_u}'{H_d}'+MH_uh_d+M'{H_d}'h_u
     +M_D\bar D^cD^c-M_D'\bar D^cd^c~.
\la{W4}
\end{eqnarray*}
For simplicity we do not couple ${D'}^c,\bar D^{'c}$ states with chiral matter and
assume that they have mass $M_{D'}\sim  M_D$.
After integrating out the $H$ and $D$-states one can easily verify that the effective
Yukawa interactions are
$$W_4^{\rm eff}=\lam_{t'}q_4u^c_4h_u+\lam_{b'}q_4d^c_4h_d+\lam_{\tau '}l_4e^c_4h_d~,$$
where:
\begin{eqnarray}
\lam_{t'}&=&\lam_{t'}^{(1)}+\lam_U\cos \ga'\\
\lam_{b'}&=&\lam_{b'}^{(1)}+\lam_D\cos \ga +\lam_D'\cos \ga_D\\
\lam_{\tau'}&=&\lam_{\tau'}^{(1)}+\lam_E\cos \ga
\end{eqnarray}
\beq \tan \ga' \simeq \fr{M_{H'}}{M'} ~,~~~~~~\tan \ga \simeq
\fr{M_H}{M}~,~~~~~~~\tan \ga_D \simeq
\fr{M_D}{{M'}_{\hs{-0.15cm}D}}~. \la{W4eff} \eeq The relevant
diagrams are shown in Fig. \ref{fig:4seesaw}. With all the mass
scales of the same order ($\simeq \La_4$) the effective
superpotential
 given above
  is valid below
the scale $\La_4$. With $\cos \ga \approx \cos \ga_D \approx \cos \ga'\approx 1$, we can
 see that the effective (derived) Yukawas
can be non-perturbative ($\approx 3$) while the original Yukawa
couplings remain perurbative;
 for example, $\lam_{t'}\simeq 2.4$ with
$\lam_{t'}^{(1)}\simeq \lam_U\simeq 1.2$. Above the scale $\La_4$ we
are dealing with the couplings $\lam_{t',b',\tau'}^{(1)}$ and
$\lam_{U,D,E}, {\lam'}_{\hs{-0.15cm}D}$.  By making proper choice
for the values of these couplings at $\La_4$, we can have a
perturbative regime up to the GUT scale. To demonstrate this we take
$\La_4=1.66\TeV$ and set up all RG equations valid above
 this scale.
At 1-loop they are given by
\begin{eqnarray}
16\pi^2\fr{d}{dt}\lam_{t'}^{(1)}&=&\lam_{t'}^{(1)}\l S_q+S_{u^c}+S_{h_u}-c_i^ug_i^2\r \\
\la{tpr1}
16\pi^2\fr{d}{dt}\lam_{b'}^{(1)}&=&\lam_{b'}^{(1)}\l S_q+S_{d^c}+S_{h_d}-c_i^dg_i^2\r \\
\la{bpr1}
16\pi^2\fr{d}{dt}\lam_{\tau'}^{(1)}&=&\lam_{\tau'}^{(1)}\l S_l+S_{h_d}-c_i^eg_i^2\r \\
\la{taupr1}
16\pi^2\fr{d}{dt}\lam_{t}&=&\lam_{t}\l 6\lam_t^2+3\left(\lam_{t'}^{(1)}\right)^2-c_i^ug_i^2\r \\
\la{t}
16\pi^2\fr{d}{dt}\lam_{U}&=&\lam_{U}\l S_q+S_{u^c}+3\lam_U^2-c_i^ug_i^2\r \\
\la{U}
16\pi^2\fr{d}{dt}\lam_{D}&=&\lam_{D}\l S_q+S_{d^c}+3\lam_D^2+\lam_E^2-c_i^dg_i^2\r \\
\la{D}
16\pi^2\fr{d}{dt}{\lam'}_{\hs{-0.15cm}D}&=&{\lam'}_{\hs{-0.15cm}D}\l S_q+S_{h_d}+2\lam_D'^2-c_i^dg_i^2\r \\
\la{D2}
16\pi^2\fr{d}{dt}\lam_{E}&=&\lam_{E}\l S_l+3\lam_D^2+\lam_E^2-c_i^eg_i^2\r
\la{E}
\end{eqnarray}
where
\begin{eqnarray}
S_q&=&\left(\lam_{t'}^{(1)}\right)^2+\left(\lam_{b'}^{(1)}\right)^2+\lam_U^2+\lam_D^2+\lam_D'^2\\
S_{u^c}&=&2\left(\lam_{t'}^{(1)}\right)^2+2\lam_U^2\\
S_{d^c}&=&2\left(\lam_{b'}^{(1)}\right)^2+2\lam_D^2\\
S_l&=&3\left(\lam_{\tau'}^{(1)}\right)^2+3\lam_E^2\\
S_{h_u}&=&3\left(\lam_{t'}^{(1)}\right)^2+3\lam_t^2\\
S_{h_d}&=&3\left(\lam_{b'}^{(1)}\right)^2+\left(\lam_{\tau'}^{(1)}\right)^2+3\lam_D'^2
\end{eqnarray}
\beq
c_i^u=\l \fr{13}{15},~ 3, ~\fr{16}{3}\r ~,~~~~c_i^d=\l \fr{7}{15},~ 3,~ \fr{16}{3}
 \r ~,~~~~c_i^e=\l \fr{9}{5},~ 3,~ 0\r ~,
\la{RFfactors} \eeq and $t=\ln \mu $. We have ignored bottom and tau
Yukawa couplings  because we still work in a low $\tan \bt $ regime.
Also the Dirac Yukawa coupling of the fourth left handed neutrino
with the `right handed' singlet $N$ is neglected, because assuming
$m_{\nu'}\simeq 50$~GeV we get  $\lam_{\nu'}\simeq 0.25$ which is small.

\begin{figure}[!ht]
  \begin{flushleft}
 \vs{0cm}
    \begin{tabular}{cc}\hs{-6.5cm}
            \resizebox{1.07\textwidth}{!}{\includegraphics{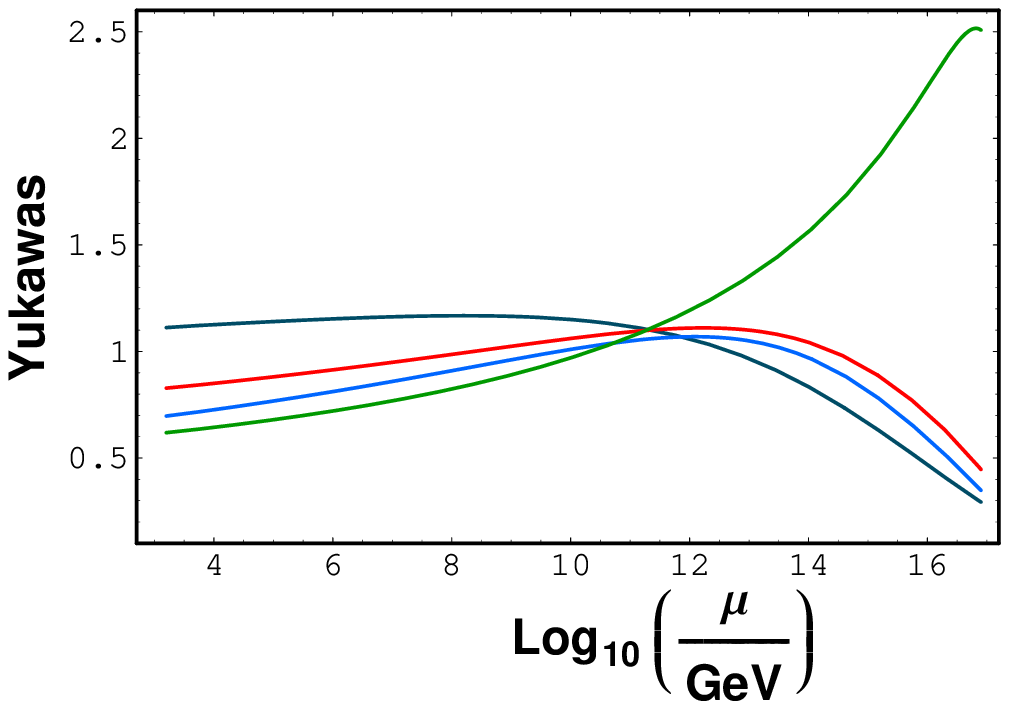}} 
            \put(-185,608){$\lam_{\tau'}^{(1)}$}
            \put(-300,578){$\lam_{t}$}
            \put(-160,560){$\lam_{b'}^{(1)}$}
\put(-270,547){$\setminus $}
            \put(-268,533){$\lam_{t'}^{(1)}$}
            &\hs{-10cm}
      \resizebox{1.07\textwidth}{!}{\includegraphics{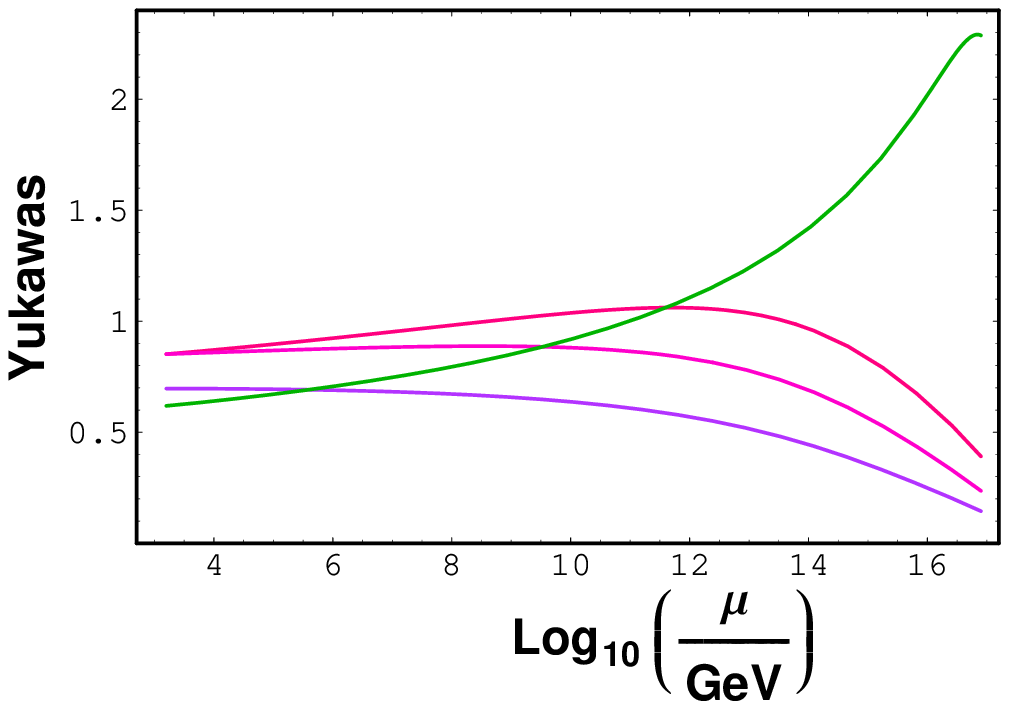}}
      \put(-185,608){$\lam_E$}
            \put(-160,560){$\lam_D$}
            \put(-190,555){${\lam'}_{\hs{-0.15cm}D}$}
           \put(-200,528){$\lam_U$}
       \\
    \end{tabular}
    \vs{-14.5cm}
    \caption{Plots at left hand side: running of Yukawa couplings $\lam_{t}$, $\lam_{t'}^{(1)}$, $\lam_{b'}^{(1)}$ and $\lam_{\tau'}^{(1)}$.
    Right hand side: running of couplings $\lam_U$, $\lam_D$, ${\lam'}_{\hs{-0.15cm}D}$, $\lam_E$.}
    \label{fig:yukRG}
  \end{flushleft}
\end{figure}

At scale $\La_4$, for  boundary conditions we take
$$
{\rm at}~~\mu =\La_4=1.66\TeV:~~~~\lam_{t'}^{(1)}=\lam_U=0.697~,
$$
\beq
\lam_{b'}^{(1)}=0.828~,~~
\lam_D={\lam'}_{\hs{-0.15cm}D}=0.852~,~~~\lam_{\tau'}^{(1)}=\lam_E=0.616~,
\la{boundY}
\eeq
and run the couplings up to $\mu =M_{\rm GUT}$. The numerical solutions are displayed
 in Fig. \ref{fig:yukRG}.
For completeness we have also included 2-loop contributions. As we
see from Fig. \ref{fig:yukRG}, all couplings remain perturbative.
Note that the boundary values in (\ref{boundY}) with $\cos \ga
\approx \cos \ga'\approx 1$
 for $\tan \bt \simeq 2$ give values for
 $m_{t'}, m_{b'} , m_{\tau'}$ (evaluated at their own mass scales) satisfying current
 experimental bounds.
Thus, our solution is fully consistent.

\begin{figure}[!ht]
   \begin{flushleft}
       \hs{3cm}\includegraphics[scale=1]{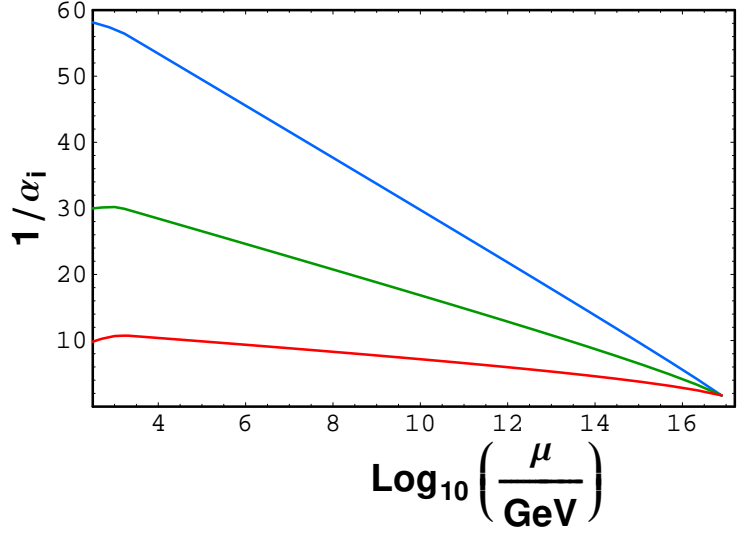}
 \vs{1cm}
        \caption{Gauge coupling unification.$M_{\rm GUT}=7.77\cdot 10^{16}\GeV $, $\fr{\al_{\rm GUT}}{4\pi }=0.046$}
        \label{fig:coupUnif}
   \end{flushleft}
\end{figure}

With this extension and values of the couplings given in (\ref{boundY}),
the gauge coupling unification occurs at relatively high scale 
$M_{\rm GUT}\simeq 7.8\cdot 10^{16}\GeV$ with perturbative unified gauge coupling
$\fr{\al_{\rm GUT}}{4\pi }\simeq 0.046$. 
The corresponding picture is shown in Fig. \ref{fig:coupUnif}.
For this case, for the mass
of   ${D'}^c,\bar D^{'c}$ states we took $M_{D'}\simeq 2.09\TeV$. In our analysis
we have included 2-loop corrections and also weak scale threshold
effects due to $m_t, m_{t'}, m_{b'}$ and $m_{\tau'}$.
 Note that the gauge coupling unification scale is somewhat
   higher compared to that of the usual SUSY GUT ($\simeq 2\cdot  10^{16}\GeV$).
   This will help to alleviate the proton decay problem in SUSY GUT.

We have demonstrated that with a simple extension one can make the MSSM with four chiral generations perturbative all the way up to the GUT scale. This gives firm ground for embedding the whole scenario in a Grand Unified Theory. Other variations of the construction of the effective Yukawa sector are possible, however, we have limited ourselves here with one example because it solves the problems in a simple and efficient way. We hope that our studies will motivate others in further investigations.

\section{Conclusions}
\label{conc}

We have investigated the implications of the presence of a 4th chiral family of fermions in the MSSM as well as the SM. The precision EW parameters $S$ and $T$ set constraints on the masses of the 4th family and the splitting between the up and down type quarks ($t'$ and $b'$).  We have plotted the allowed regime of 4th generation quark masses in Fig. (\ref{fig:both}).

We also investigated the constraint on the 4th family from the perturbativity condition on the  corresponding Yukawa couplings, and found that in MSSM, there is no allowed value of $\tan \bt $ for which the couplings remain perturbative all the way up to the GUT scale. As a result, if a 4th family is discovered at the LHC, then for the theory to make sense perturbatively, there must be additional new physics with a suitable ultraviolet completion. We have presented such a model with additional vector-like states, at the TeV scale. In our model, only the very narrow range of $\tan \bt < 2$ is allowed. 

In addition to observing the 4th chiral family of fermions at the LHC, the model has several predictions, such as the existence of vector-like down type quarks at the TeV scale which can be pair produced by gluon-gluon fusion, enhanced decay of the lightest Higgs boson to two photons, and enhanced Higgs production from gluon-gluon fusion due to the $t'$ and $b'$ quarks. These predictions of the model can be tested at the LHC.

\section*{\small Acknowledgement}
\vs{-0.2cm} We are grateful to K.S. Babu for useful discussions and
comments. We also thank G. Kribs for his useful comments on the previous version of this paper.
SN and ZM would like to thank the Theoretical Physics Department
of Fermilab for warm hospitality and support during the completion
of this work. The work is supported in part by DOE grant
DE-FG02-04ER41306 and DE-FG02-ER46140. Z.T. is also partially
supported by GNSF grant 07\_462\_4-270.

\newpage

\newpage

\newpage

\end{document}